\documentclass[aps,prl,floatfix,twocolumn]{revtex4}

\usepackage{amsmath,units,xspace}
\newcommand{\micron}{\ensuremath{\unit{\mu m}}\xspace}

\usepackage{graphicx}
\graphicspath{{figures/}}

\begin{document}

\title{Holographic assembly of quasicrystalline
  photonic heterostructures}

\author{Yael Roichman}
\author{David G. Grier}

\affiliation{Department of Physics and Center for Soft Matter Research,
  New York University, New York, NY 10003}

\date{May 31, 2005}


\begin{abstract}
Quasicrystals have a higher degree of 
rotational and point-reflection symmetry than conventional 
crystals.
As a result, quasicrystalline heterostructures fabricated
from dielectric materials with micrometer-scale features
exhibit interesting and useful optical properties including
large photonic bandgaps in two-dimensional systems.
We demonstrate the holographic assembly of two-dimensional
and three-dimensional dielectric quasicrystalline heterostructures,
including structures with specifically engineered defects.
The highly uniform quasiperiodic arrays of optical traps used
in this process also provide model aperiodic 
potential energy landscapes for fundamental studies of
transport and phase transitions in soft condensed matter systems.
\end{abstract}



\maketitle

\begin{figure}[b]
  \centering
  \includegraphics[width=0.8\columnwidth]{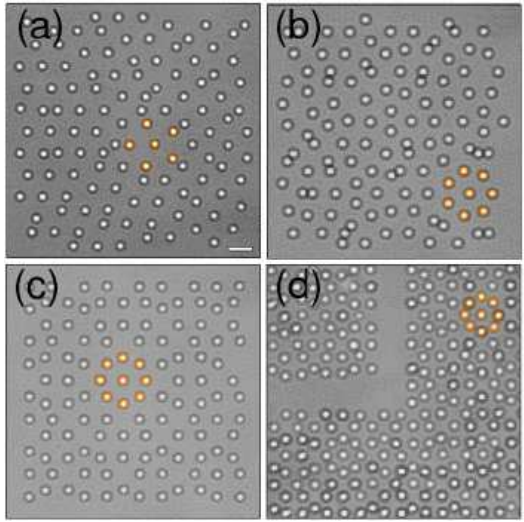}
  \caption{Two-dimensional colloidal quasicrystals organized with
    holographic optical traps.  (a) 5-fold.  (b) 7-fold. (c) 8-fold.
    (d) An octagonal quasicrystal with an embedded waveguide.  Scale
    bar indicates 5~\micron.}
  \label{fig:2d}
\end{figure}

Quasicrystals have long-ranged orientational order even though they 
lack the translational periodicity of crystals.
Not limited by conventional spatial point groups, they can adopt 
rotational symmetries that are forbidden to crystals.
The resulting large number of effective reciprocal lattice vectors 
endows quasicrystals' effective Brillouin zones with an unusually
high degree of rotational and point inversion symmetry \cite{burkov92}.
These symmetries, in turn, facilitate the
the development of photonic band gaps (PBG)
\cite{joannopoulos95} for light propagating through 
quasicrystalline dielectric
heterostructures \cite{chan98,cheng99b,Zhang01}, even when the
dielectric contrast among the constituent materials is low.
Photonic band gaps have been realized in one- \cite{hattori94}
and two-dimensional \cite{zoorob00}
lithographically defined quasiperiodic structures.
Here we demonstrate rapid assembly of arbitrary materials into 
two- and three-dimensional quasicrystalline
heterostructures with features suitable for photonic device applications.

\begin{figure}[t]
  \centering
  \includegraphics[width=0.8\columnwidth]{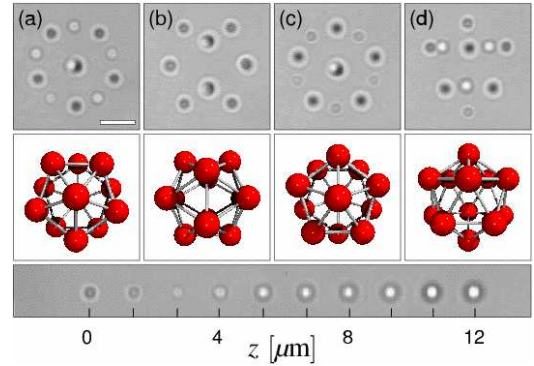}
  \caption{Four views of a rolling colloidal icosahedron. (a) 5-fold
  axis. (b) 2-fold axis. (c) 5-fold axis. (d) Midplane.  Scale
  bar indicates 5~\micron.  The complete assembly and rolling process
  is shown in the associated movie.}
  \label{fig:icosahedron}
\end{figure}

Our approach is based on the holographic optical trapping technique
\cite{dufresne98,grier03,polin05} in which computer-generated
holograms are projected through a high-numerical-aperture microscope
objective lens to create large three-dimensional arrays of optical traps.
In our implementation, light at 532~\unit{nm}
from a frequency-doubled diode-pumped
solid state laser (Coherent Verdi) is imprinted with phase-only
holograms using a liquid crystal spatial light modulator (SLM)
(Hamamatsu X8267 PPM).  The modified laser beam is relayed to
the input pupil of a $100\times$ NA 1.4 SPlan Apo oil immersion
objective mounted in an inverted optical microscope (Nikon TE-2000U),
which focuses it into traps.
The same objective lens is used to form images of trapped objects,
using the microscope's conventional imaging train \cite{polin05}.

We used this system to organize colloidal silica microspheres
1.53~\micron in diameter (Duke Scientific Lot 5238) dispersed in
an aqueous solution of $180:12:1$ (wt/wt) acrylamide, 
$N,N^\prime$-methylenebisacrylamide and diethoxyacetophenone (all Aldrich electrophoresis grade).  
This solution rapidly
photopolymerizes into a transparent
polyacrylamide hydrogel under ultraviolet illumination, and is stable otherwise.
Fluid dispersions were imbibed into
30~\micron thick slit pores formed by bonding the edges of
\#1 coverslips to the faces of glass microscope slides.
The sealed samples were then
mounted on the microscope's stage for processing and analysis.

\begin{figure}[t]
  \centering
  \includegraphics[width=0.8\columnwidth]{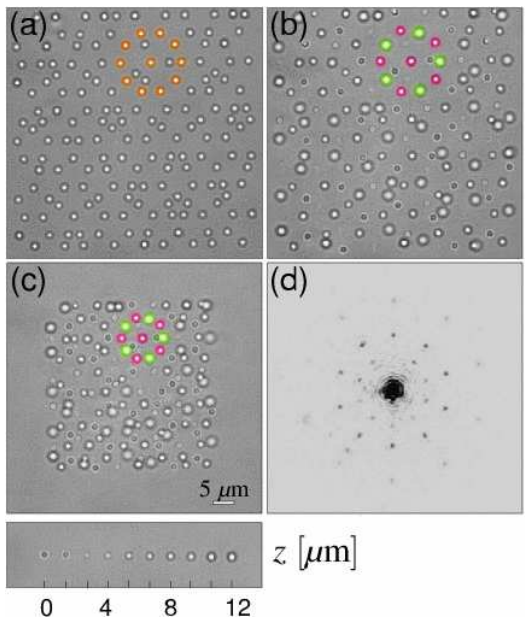}
  \caption{Holographic assembly of a three-dimensional colloidal
  quasicrystal.  (a) Colloidal particles trapped in a two-dimensional
  projection of a three-dimensional icosahedral quasicrystalline
  lattice.  (b) Particles displaced into the fully three-dimensional
  configuration.  The shaded region identifies one embedded
  icosahedron.  (c) Reducing the lattice constant creates a compact
  three-dimensional quasicrystal.  (d) Optical diffraction pattern
  showing ten-fold symmetric peaks.  The three-dimensional assembly
  process is shown in the associated movie.}
  \label{fig:3d}
\end{figure}

Silica spheres are roughly twice as dense as water and sediment
rapidly into a monolayer above
the coverslip.
A dilute layer of spheres is readily organized by holographic optical tweezers
into arbitrary two-dimensional
configurations, including the quasicrystalline examples in Fig.~\ref{fig:2d}.
Figures~\ref{fig:2d}(a), (b) and (c) show planar pentagonal,
heptagonal and octagonal quasicrystalline domains \cite{suck}, respectively,
each consisting of more than 100 particles.
Highlighted spheres emphasize each domain's symmetry.
These structures all have been
shown to act as two-dimensional PBG materials in
microfabricated arrays of posts and holes
\cite{chan98,jin99,bayindir01,escuti04}. 
As a soft fabrication technique,
holographic assembly requires substantially less processing
than conventional methods such as electron-beam lithography, and
can be applied to a wider range of materials. 
Unlike complementary optical
fabrication techniques such as multiple-beam holographic photopolymerization
\cite{escuti04,gauthier04,gauthier05,Wang03},
assembly with holographic optical traps lends itself to creating
nonuniform architectures with specifically engineered 
features, such as the channel 
embedded in the octagonal domain in
Fig.~\ref{fig:2d}(d).  Similar structures of comparable dimensions 
have been shown to
act as narrow-band waveguides and frequency-selective
filters for visible light \cite{jin99,bayindir01,chen99,jin00}.

Holographic trapping's ability to assemble free-form heterostructures 
extends also to three dimensions.
The sequence of images of a rolling
icosahedron in Fig.~\ref{fig:icosahedron} shows how
the colloidal spheres' appearance changes with distance from
the focal plane.
This sequence also recalls earlier reports \cite{leach04,sinclair04}
that holographic traps can successfully organize spheres into vertical
stacks along the optical axis, while maintaining one sphere in each
trap.  

The icosahedron itself is the fundamental building block of
a class of three-dimensional quasicrystals, such as the example in
Fig.~\ref{fig:3d}. Building upon our earlier work on holographic
assembly \cite{korda02}, we assemble a three-dimensional
quasicrystalline
domain by first creating a two-dimensional
arrangement of spheres corresponding to the planar projection of
the planned quasicrystalline domain, Fig.~\ref{fig:3d}(a).  Next,
we translate the spheres along the optical axis
to their final three-dimensional coordinates in the
quasicrystalline domain, as shown in Fig.~\ref{fig:3d}(b).  
One icosahedral unit is highlighted in
Figs.~\ref{fig:3d}(a) and (b) to clarify this process.
Finally, the separation between the traps is decreased
in Fig.~\ref{fig:3d}(c) to create an optically dense
structure. This particular domain consists of 173 spheres in
roughly 7 layers, with typical inter-particle separations of
3~\micron.

The completed quasicrystal was gelled and its optical diffraction pattern
recorded at a wavelength of 632~\unit{nm} by illuminating the
sample with a collimated beam from a HeNe laser, collecting the
diffracted light with the microscope's objective lens and projecting
it onto a charge-coupled device (CCD) camera with a Bertrand lens.
The well-defined diffraction spots clearly reflect the quasicrystal's
five-fold rotational symmetry in the projected plane.

Holographic assembly of colloidal silica quasicrystals in water is
easily generalized to other materials and solvents.
Deterministic organization of disparate components under holographic control
can be used to embed gain media in
PBG cavities, to install materials with nonlinear optical properties within 
waveguides to form switches,
and to create domains with distinct chemical functionalization.
The comparatively small domains we have created can be combined into
larger heterostructures through sequential assembly and spatially
localized photopolymerization.
In all cases, this soft fabrication process results in mechanically
and environmentally stable materials that can be integrated readily into
larger systems.

Beyond the immediate application of holographic trapping to
fabricating quasicrystalline materials,
the ability to create and continuously optimize
such structures provides new opportunities for studying the
dynamics \cite{polin05} and statistical mechanics \cite{denton98} 
of colloidal quasicrystals.
The optically generated quasiperiodic potential energy landscapes
developed for this study 
also should provide a flexible model system for experimental studies
of transport \cite{korda02b} through aperiodically
modulated environments.

We are grateful to Paul Steinhardt, Paul Chaikin and Weining
   Man for illuminating conversations.  Support was provided by the
   National Science Foundation through Grant Number DMR-0451589.

\end{document}